\documentclass[journal,onecolumn, draftcls, 12pt]{IEEEtran} 

\usepackage[utf8]{inputenc}
\usepackage[T1]{fontenc}
\usepackage[hidelinks]{hyperref}
\usepackage{url}
\usepackage{ifthen}
\usepackage{cite}
\usepackage[cmex10]{amsmath} 
\usepackage{amsfonts,amssymb}
\usepackage{color}
\usepackage{ifthen}
\usepackage[arxiv]{optional} 
\usepackage{accents}
\newlength{\dhatheight}

\DeclareMathOperator{\E}{\mathbb{E}}
\DeclareMathOperator{\va}{\mathbf{a}}
\DeclareMathOperator{\vb}{\mathbf{b}}
\DeclareMathOperator{\vc}{\mathbf{c}}
\DeclareMathOperator{\vd}{\mathbf{d}}
\DeclareMathOperator{\vu}{\mathbf{u}}
\DeclareMathOperator{\vp}{\mathbf{p}}
\DeclareMathOperator{\vj}{\mathbf{j}}
\DeclareMathOperator{\vl}{\mathbf{l}}

\begin{document}
\title{A dynamic program for linear sequential coding for Gaussian MAC with noisy feedback}

 \author{%
  \IEEEauthorblockN{Deepanshu Vasal}\\
  \IEEEauthorblockA{Northwestern University\\
                     Evanston, IL, 60201 USA\\
                     \texttt{dvasal@umich.edu}}
 }

\maketitle


\begin{abstract}
In this paper consider a two user multiple access channel with noisy feedback. There are two senders with independent messages who transmit symbols across an additive white Gaussian channel to a receiver, who in turn sends back a symbol which is received by the two senders through two independent noisy Gaussian channels. We consider the case when the feedback is active i.e. the receiver actively encodes the feedback using a linear state process. We pose this as a problem of linear sequential coding at the senders and the receiver to minimize the terminal mean square probability of error at the receiver. This is an instance of decentralized control with no common information at the senders and the receiver. In this paper, we construct two linear
controllers at the sender and the receiver. Due to linearity of the
policies and the controllers, all the random variables involved are
jointly Gaussian. Moreover, the corresponding covariance matrix
at the receiver of the estimation process of the senders' messages is a deterministic process,
which is a function of the parameters of the controllers and the
strategies of the players, and is thus perfectly observed by the
senders. Based on this observation, we use deterministic dynamic
programming to find the optimal policies and the optimal linear
controllers at both the senders and the receiver. The problem with passive feedback can be considered as a special case.
\end{abstract}


\def\cE{\mathcal{E}}
\def\cX{\mathcal{X}}
\def\cY{\mathcal{Y}}
\def\cZ{\mathcal{Z}}
\def\cW{\mathcal{W}}
\def\cP{\mathcal{P}}
\def\cU{\mathcal{U}}
\def\cV{\mathcal{V}}
\def\cR{\mathcal{R}}
\def\cC{\mathcal{C}}
\def\cS{\mathcal{S}}
\def\cF{\mathcal{F}}
\def\cG{\mathcal{G}}
\def\cB{\mathcal{B}}

\def\tw{\tilde{w}}
\newcommand{\eq}[1]{\begin{align}#1\end{align}}
\newcommand{\seq}[1]{\begin{subequations}#1\end{subequations}}
\newcommand{\lb}[1]{\left\{ \begin{array}{ll} #1 \end{array} \right.}
\newcommand\dif{\mathop{}\!\mathrm{d}}
\newcommand{\bm}[1]{\begin{bmatrix}#1\end{bmatrix}}
\newcommand{\bit}[1]{\begin{itemize}#1\end{itemize}}

\newcommand{\cA}{\mathcal{A}}

\newcommand{\cH}{\mathcal{H}}

\newcommand{\tcC}{\tilde{\mathcal{C}}}
\newcommand{\hV}{\hat{V}}
\newcommand{\hu}{\hat{u}}
\newcommand{\hN}{\hat{N}}
\newcommand{\tx}{\tilde{x}}
\newcommand{\hz}{\hat{z}}
\newcommand{\hw}{\hat{w}}
\newcommand{\ha}{a^2}
\newcommand{\hx}{\hat{x}}
\newcommand{\hcC}{\hat{\mathcal{C}}}
\newcommand{\hD}{d^2}
\newcommand{\cN}{\mathcal{N}}
\newcommand{\tgamma}{\tilde{\phi}}
\newcommand{\Qx}[1]{Q^{i}(#1)}
\newcommand{\Qw}[1]{Q_w^{i}(#1)}
\newcommand{\defeq}{\buildrel\triangle\over =}
\newcommand{\apos}{\textsc{\char13}}
\newcommand{\pushright}[1]{\ifmeasuring@ #1 \else\omit\hfill$\displaystyle#1$\fi\ignorespaces}
\newcommand{\pushleft}[1]{\ifmeasuring@ #1 \else\omit$\displaystyle#1$\hfill\fi\ignorespaces}
\newcommand{\nn}{\nonumber}
\newcommand{\rd}{\right.}
\newcommand{\ld}{\left.}

\newtheorem{lemma}{Lemma}
\newtheorem{fact}{Fact}
\newtheorem{theorem}{Theorem}
\newtheorem{assumption}{Assumption}

\newcommand{\eqdef}{\stackrel{\scriptscriptstyle \triangle}{=}}
\newcommand{\mdef}{\stackrel{\text{\tiny def}}{=}}
\def\Real{\mathbb{R}}
\def\P{\mathbb{P}}


\section{Introduction}

The fundamentals of digital communication were laid down by Shannon in his pioneering work in~\cite{Sh48}. Since then there has been significant effort on finding coding schemes that minimize probability of error and achieve capacity. There has been a lot of focus on a point to point discrete memoryless channel for which efficient codes such as Turbo codes, LDPC and polar codes have been formulated. For a point to point channel with feedback, it is known that the feedback doesn't increase the capacity~\cite{Sh56} but it can significantly improve the error exponents from exponential to super exponential~\cite{ScKa66}. This can drastically reduce the delay incurred by real time communication such as in cellphones and video calls, reduce coding and decoding complexity and sequential coding schemes can be of arbitrary length, and thus quite appropriate for real time communication.

There have been multiple works proposing transmission schemes for several instances of channel with \emph{noiseless} feedback such as Horstein's scheme~\cite{Ho63} for binary symmetric channel (BSC), and Schalkwijk and Kailath~\cite{ScKa66} for an additive white Gaussian noise (AWGN) channel, all of which were generalized by a posterior matching scheme (PMS)~\cite{ShFe08} for an arbitrary channel. However, it is known that these schemes perform rather poorly when the feedback is even slightly noisy~\cite{Sc66}. The problem of finding optimum transmission schemes for noisy feedback has been an important open problem in information theory. In this problem, both the sender and the receiver receive different observations whose domain increase exponentially in time, and the set of possible strategies grow double exponential in time. Because of asymmetry of information and lack of any common information, there is no known (dynamic programming like) methodology that decomposes this problem in time reducing the complexity to linear in time. It was shown in~\cite{Jietal19} that a scheme using RNN (recurrent neural networks) improves the current best known scheme by three orders of magnitude. 

Ozarow considered multiple access channel(MAC) with \emph{noiseless} feedback in~\cite{Oz84} proposed a linear scheme for this channel that achieves capacity. However, capacity of MAC with \emph{noisy} feedback is still unknown. The
capacity region of the general two channel is still
unknown. Recently,~\cite{Va20state} proposed a sequential decomposition methodology to decompose point to point channel with noisy feedback across time. This scheme was specialized for Gaussian channel with \emph{passive} feedback in~\cite{MiVaKi20} to find a linear sequential scheme, and that idea has been used to provide linear strategies using a dynamic programming methodology in various multi user Gaussian settings~\cite{ Va21a,Va21b,Va21d,Va21e,Va21f}.
In this paper, we consider MAC with \emph{noisy} feedback. Based on the idea of ``auxiliary controller" introduced in~\cite{Va20state}, we propose a dynamic program for MAC with noisy active feedback, where both the senders as well as the receiver actively codes at each time instant. 
More specifically, we construct linear controllers at the two senders as well as the receiver, where the receiver estimates the senders' messages through its observations. We show that due to linearity of the policies and the controllers, and due to Gaussianity of the random processes involved, the co variance matrices at the receiver of the senders' messages are deterministic processes which are function of the parameters of the controllers and the strategies of all the users. Based on this observation, we use dynamic programming with state as the covariance matrices to find the optimal policies of all the three users, where the terminal cost is given by error variance of the messages at the receiver.

\section{Channel Model}
\label{sec:model}

We consider an additive Gaussian multiple access channel (MAC) with \emph{active} noisy feedback.
Consider the problem of transmission of messages $m^1\in\{1,2,\ldots \mathbb{M}^1 \}$ at sender 1 and $m^2\in\{1,2,\ldots \mathbb{M}^2 \}$ at sender 2, over the MAC Gaussian channel with noisy feedback using fixed length codes of length $T$. At each instant $t$, the sender $i$ transmits a symbol $x_t^i\in\cX^i=\mathbb{R}$, while the receiver observes $y_t^r\in\cY=\mathbb{R}$, where $y_t^r = x_t^1 + x_t^2 + w_t^f$ and $w^f_t = N(0,(\sigma^f)^2)$. Then the receiver sends back $x_t^r$ to the senders through different independent additive Gaussian noisy channels such that sender 1 receives $y_t^1 = x_t^r+w_t^{b,1}$ and sender 2 receives $y_t^2 = x_t^r+w_t^{b,2}$ (Note that the reverse channel is a broadcast channel).
We assume that 
the messages $m^i\in\{1,2,\ldots \mathbb{M}^i \}$ are uniformly distributed and mutually independent,
the forward channel $Q^f$ and is an independent Gaussian channel with 0 mean and known variance, $(\sigma^{f})^2$. Similarly the backward channels are independent Gaussian channels with 0 mean and variances $(\sigma^{b,i})^2$. Inspired by~\cite{Va20state}, we assume that both the senders and the receiver individually maintain controllers $\vu_t^{1},\vu_t^{2}, \vu_t^r$, respectively, where $\vu_t^{i}\in\mathbb{R}^3$, is a controlled process maintained by the sender $i$ at her end. Similarly, the receiver maintains $\vu_t^{r} \in\mathbb{R}^3$. We assume $\vu_t^{1},\vu_t^{2},\vu_t^r$ are linearly updated such that 
\eq{
\vu_{t+1}^{1} &= \va_t^{1}\vu_t^{1} + \vb_t^1m^1+ \vc_t^{1} y^1_t,\\
\vu_{t+1}^{2} &= \va_t^{2}\vu_t^{2} + \vb_t^2m^2+ \vc_t^{2} y^2_t\\
\vu_{t+1}^r &= \va^r_t\vu^r_t + \vc_t^ry^r_{t},
} where $\va_t^{i}\in\mathbb{R}^9,\vb_t^i \in \mathbb{R}^3,\vc_t^{i}\in\mathbb{R}^3$. Let $G_t^{i} := (\va_t^{i},\vb_t^i,\vc_t^{i})$.

At each instant $t$, the sender $i$ generates its channel inputs $x_t^i$ based on its private message $m^i$ and $\vu_t^{i}$, where as mentioned above, $\vu_t^{i}$ is a function of the noisy feedback $y^i_{1:t-1}$. We assume linear strategies of the players of the form $x_t^i = \phi_t^i(m^i,\vu_t^i) = k_t^im^i+\vd^i_t \vu_t^i$. Without loss of generality, we assume $k^i_t = 0$. 
For each time $t$, we assume $var(X_t^i) = P^i_t$, for all $i,t$, where we consider two cases (i) Instantaneous power constraint: $P_t^i \leq P^i/T, \ \forall i,t$, and (ii) Total power constraint: $\sum_{t=1}^T P_t^i \leq P^i,\ \forall i$. $var(X_t^i) = P^i_t$  implies 
\eq{
    \vd_t^1&= \frac{\sqrt{P_t^1}}{\sqrt{3}}\bm{1&1& 1} (\Sigma_{\vu_{t}^1})^{-1/2}\\
    \vd_t^2&= \frac{\sqrt{P_t^1}}{\sqrt{3}}\bm{1&1& 1}(\Sigma_{\vu_{t}^2})^{-1/2}.
}.

For each time $t$, the decoder estimates the messages $m^1, m^2$ based on $t$ channel outputs, $y^1_{1:t}$ as
\begin{equation}
(\hat{m}^1_t,\hat{m}^2_t) = g(y^1_{1:t}).
\end{equation}

A fixed-length transmission scheme for the channel is $s=(\phi^1,\phi^2,\phi^3,G^1,G^2,G^3,g)$, consisting of the encoding functions
$\phi^1,\phi^2,\phi^3,$ the update functions $G^1,G^2,G^3$ and the decoding function $g$.
The error probability associated with the transmission scheme $s$ is defined as
\begin{equation}
P_e(s) = \E^s[(M^1- \hat{M}^1)^2+(M^2- \hat{M}^2)^2].
\end{equation}

\section{Decentralized control of Gaussian MAC with active noisy feedback}
\label{sec:dsaht}

One may pose the following optimization problem.
Given the alphabets $\mathcal{M}^1,\mathcal{M}^2,\cX$, $\cY$, $\cZ$, the channels $Q^f$, and for a fixed length $T$, design the optimal transmission scheme $s=(\phi^1,\phi^2,G,g)$ that minimizes the error probability $P_e(s)$.
\begin{equation}
P_e = \min_s \P_e(s) \tag{\textbf{P1}}
\end{equation}


For any pair of encoding and update functions, the optimal decoder is the ML decoder (assuming equally likely hypotheses), denoted by $g_{ML}$.
%

\subsection{Unconditioned variances at the senders and the receiver}
We know that
\begin{align}
    \vu^1_{t+1} &= \va_t^1\vu^1_t+\vb_t^1m^1+\vc^1_ty^1_t\label{Eqn:ut_transition}\\
    \vu^2_{t+1} &= \va_t^2\vu^2_t+\vb_t^2m^2+\vc^2_ty^2_t\\
    \vu^r_{t+1} &= \va_t^r\vu^r_t+\vc^r_ty^r_t\label{Eqn:ut_transition}\\
    y_t^1&= x_t^r + w_t^{b,1}\\
    y_t^2&= x_t^r + w_t^{b,2}\\
    y_t^r&= x_t^1 + x_t^2 + w_t^{f}\\
    x_{t}^1 &= \vd_t^1 \vu^1_{t}, \label{Eqn:power_normalized}\\
    x_{t}^2 &=  \vd_t^2 \vu^2_{t},\\
    x_t^r &=  \vd_t^r \vu^r_{t}
\end{align}


The power of variable $\vu_t^i$ at each instant can be represented as a function of the power at the previous instant. From~\eqref{Eqn:ut_transition}, we have,
\begin{align}
 \vu^1_{t+1} &= \va_t^1\vu^1_t+\vb_t^1m^1 + \vc^1_tx^r_t +\vc_t^1 w_t^{b,1}\\
     			    & = \va_t^1\vu^1_t  +\vb_t^1m^1 + \vc^1_t\vd_t^r\vu^r_t + \vc^1_t w_t^{b,1} \\
     			    \vu^2_{t+1} &= \va_t^2\vu^2_t +\vb_t^2m^2 + \vc^2_tx_t^r+\vc_t^2 w_t^{b,2}\\
     			    & = \va_t^2\vu^1_t  +\vb_t^2m^2+ \vc^2_t\vd_t^r\vu^r_t
     			    + \vc^2_t w_t^{b,2}\\
     			    \vu_{t+1}^r &= \va^r_t\vu^r_t + \vc_t^ry^r_{t}\\
     		&=\va^r_t\vu^r_t + \vc_t^r(\vd_t^1 \vu^1_{t} + \vd_t^2 \vu^2_{t} + w_t^{f})
\end{align} 
Let $q_t:=\bm{m^1& m^2& \vu_{t}^1& \vu_{t}^2& \vu_t^r}'$. Then
using the above equations, one can write 
\eq{
\bm{m^1\\ m^2\\\vu_{t+1}^1\\ \vu_{t+1}^2\\ \vu_{t+1}^r} &= \va_t^q
\bm{m^1\\m^2\\ \vu_{t}^1\\ \vu_{t}^2\\\vu_t^r} + \vj_t^q\bm{ w_t^{b,1}\\w_t^{b,2}\\w_t^f},
}
where 
\eq{
&\va_t^q= \bm{1 & 0 & 0 & 0 & 0 \\
0&1 & 0 & 0  & 0\\
\vb_t^1&0 & \va_t^1&0&\vc_t^1\vd_t^r\\
0& 0&  \vc_t^r\vd_t^1& \vc_t^r\vd_t^2&  \va_t^r
},
&\vj_t^q = \bm{0&0&0\\0&0&0\\\vc_t^1&0&0\\0&\vc_t^2&0\\0&0&\vc_t^r}
}
where $\vd_t^i\Sigma_{\vu_t}(\vd_t^i)^T = P_t^i$ (assuming $k_t^i = 0$). Thus substitute $\vd_t^i$ in $\va_t^q$ as
\eq{
\vd_t^1&= \frac{\sqrt{P_t^1}}{\sqrt{3}}\bm{1&1& 1}(\Sigma_{\vu^1_{t}})^{-1/2}\\
\vd_t^2&= \frac{\sqrt{P_t^2}}{\sqrt{3}}\bm{1&1& 1}(\Sigma_{\vu^2_{t}})^{-1/2}.
}

Thus covariance matrix $\Sigma^q_t$ is updated as
\eq{
\Sigma^q_{t+1} &= \va_t^q\Sigma^q_t(\va_t^q)' + \vj_t^q\bm{\sigma^2_{b,1}&0&0\\
0&\sigma^2_{b,2}&0\\
0&0&\sigma^2_{f}}(\vj_t^q)'
}

We initiate the transmission with $\vu^1_0=\bm{m^1&0&0}, \vu^2_0 = \bm{m^2&0&0}, k^1_0 = 1, k_0^2 = 0$ such that 
\eq{
\Sigma^q_0(1,1) &= \Sigma^q_0(3,3)= \sigma_{m^1}^2,\\ \Sigma^q_0(2,2) &=\Sigma^q_0(6,6)= \sigma_{m^2}^2  
}
and all other entries of $\Sigma^q_0$ being 0. 

\subsection{Receiver's observed process}
Conditioned on its information at time $t$, the receiver faces the following linear estimation problem. Let $p^r_t = \bm{m^1&m^2 &\vu^1_t&\vu^2_t&\vu_t^r&y_t^r}'$. Then,  
\eq{
\vu^1_{t+1} &= \va_t^1\vu^1_t+\vb_t^1m^1 + \vc^1_tx^r_t +\vc_t^1 w_t^{b,1}\\
     			    & = \va_t^1\vu^1_t  +\vb_t^1m^1 + \vc^1_t\vd_t^r\vu^r_t + \vc^1_t w_t^{b,1} \\
     			    \vu^2_{t+1} &= \va_t^2\vu^2_t +\vb_t^2m^2 + \vc^2_tx_t^r+\vc_t^2 w_t^{b,2}\\
     			    & = \va_t^2\vu^1_t  +\vb_t^2m^2+ \vc^2_t\vd_t^r\vu^r_t
     			    + \vc^2_t w_t^{b,2}\\
     			    \vu_{t+1}^r &= \va^r_t\vu^r_t + \vc_t^ry^r_{t+1}\\
     		%
%
y^r_{t+1} &= x_{t+1}^1 + x_{t+1}^2 + w_{t+1}^f\\
&= \vd^1_t\vu^1_{t+1} + \vd^2_t\vu^2_{t+1}+ w_{t+1}^f
%
}
The above equations can be written as
\eq{
 \bm{m^1\\m^2\\\vu^1_{t+1} \\ \vu^2_{t+1}\\\vu^r_{t+1}\\y_{t+1}^r}&=  
 \bm{1&0&0&0&0&0\\
 0&1&0&0&0&0\\
 \vb_t^1&0&\va_t^1&0 &\vc_t^1\vd_t^r&0\\
 0&\vb_t^2&\va_t^2&0&\vc_t^2\vd_t^r&0\\
 \vc_t^r\vd_t^1\vb_t^1&\vc_t^r\vd_t^2\vb_t^2&\vc_t^r\vd_t^1\va_t^1+ \vd_t^2\va_t^2&0 &\va_t^r+\vc_t^r(\vd_t^1\vc_t^1+\vd_t^2\vc_t^2)\vd_t^r &0\\
 \vd_t^1\vb_t^1&\vd_t^2\vb_t^2&\vd_t^1\va_t^1+ \vd_t^2\va_t^2&0 &(\vd_t^1\vc_t^1+\vd_t^2\vc_t^2)\vd_t^r &0}
   \bm{m^1\\m^2\\\vu^1_{t} \\ \vu^2_t\\\vu_t^r\\y_{t}^r} \nn\\
 &+\bm{0 & 0 & 0\\
 0 & 0 & 0\\
 \vc_t^1 & 0 & 0\\
 0 & \vc_t^2 & 0\\
 0 & 0 & \vc_t^r\\
 0 & 0 & 1} \bm{w_t^{b,1}\\w_t^{b,2}\\w^f_{t+1}}\\
\vp^2_{t+1} &= \va^2_t \vp^2_t + \vj^2_tw^f_{t+1}\\
y_t^r &= \vc_t^2p_t^2
}

where $\va^r_t$, $\vj^r_t$ are defined from above and $\vc^r_t = \bm{0&0&0&0&1}$.
It is easy to note that under these conditions and from~\cite[Ch. 7, Thm 2.21]{KuVa86}, $\vp_t^r$ becomes a Gaussian random vector where its mean $\hat{p}^r_{t|t} :=E[p_{t}^r|y^r_{1:t}]$ and covariance $ \Sigma^r_{t|t}:= E[(p^r_{t}-E[p^r_{t}|y_{1:t}^r])(p^r_{t}-E[p^r_{t}|y^r_{1:t}])'|y^r_{1:t}]$ can be updated as follows

\seq{
\eq{
 \hat{p}^r_{t+1|t+1} &= \va^r_t\hat{p}^r_{t|t} + \vl^r_{t+1}(y_{t+1}^r- \vc^r_{t+1}\va^r_t\hat{p}^r_{t|t})\label{eq:mean_update}\\
 &= (I - \vl^r_{t+1}\vc^r_{t+1})\va^r_t\hat{p}^r_{t|t}+ \vl^r_{t+1}y^r_{t+1}\\
 \hat{p}^r_{0|0} &= \vl_0^r y^r_0\\
 \Sigma^r_{t+1|t+1}&= (I-\vl^r_{t+1}\vc^r_{t+1})\Sigma^r_{t+1|t}\\
 \Sigma^r_{t+1|t}&=
 \left(\va^r_t\Sigma^r_{t|t}\va_t^{r,'}+(\sigma^f)^2 \vj^r_t\vj^{2,'}_t\right),\\
 \Sigma^r_{0|0} &= (I-\vl^r_{0}\vc^r_{0})\Sigma^r_{0}\\
\label{eq:cov_update}
}
\label{eq:meancov_update}
where
\eq{\vl^r_{t+1} &= \Sigma^r_{t+1|t}\vc_{t+1}^{r,'}[\vc^r_{t+1}\Sigma^r_{t+1|t}\vc_{t+1}^{r,'}]^{-1}\\ \label{eq:g_def}
\vl^2_0 &= \Sigma^r_{0}\vc_0^{r,'}[\vc^r_0\Sigma^r_{0}\vc_0^{r,'}]^{-1}
}
}
and 
\eq{
\Sigma^r_0(1,1) &= \Sigma^r_0(3,3)= \sigma_{m^1}^2,\\ \Sigma^r_0(2,2) &=\Sigma^q_0(6,6)= \sigma_{m^2}^2  
}
and all other entries of $\Sigma^r_0$ being 0.

\section{A dynamic programming approach}
We note that the conditional covariance process $\Sigma^2_{t|t}$ at the receiver and the unconditional covariance process $\Sigma_t^q$ are deterministic (and thus observable) processes, controlled by the linear strategies $\phi^i$ (and equivalently $\vd_t^i$) and update functions $G^i_t$ (and equivalently $\va^i_t,\vb_t^i\vc^i_t$).

\subsection{A dynamic program for instantaneous power constraint}
In this section, we assume that both the senders and the receiver have instantaneous power constraints such that $var(X_t^i)=P_t^i\leq P^i/T,\forall i,t$. Instead of explicitly optimizing on policy parameters $P_t^i$, we force $P_t^i = P^i/T$, such that at each instant, both sender and the receiver transmit at their maximum available power.

Let $(\Sigma^q_{t+1},\Sigma^1_{t+1|t+1},\Sigma^2_{t+1|t+1},\Sigma^r_{t+1|t+1}) = \tau^a(\Sigma^q_{t},\Sigma^1_{t|t},\Sigma^2_{t|t},\Sigma^r_{t|t},G_t^1,G_t^2,G_t^3)$, where $\tau^a$ is defined through previous section.
Based on this, we propose a \emph{deterministic dynamic program} to compute optimal linear policies as follows.
\begin{enumerate}
    \item $\forall$ $\Sigma^q_{T}$, $\Sigma^1_{T},\Sigma^2_{T},\Sigma^r_{T}$,
            $  V_{T}({\Sigma^q_{T},\Sigma^1_{T},\Sigma^2_{T},\Sigma^r_{T}}) := (\Sigma^r_T(1,1))^2 + (\Sigma^2_{T}(2,2))^2 $.
    \item For $t=T-1,\ldots,1$, $\forall$ ${\Sigma^q_{T},\Sigma^1_{T},\Sigma^2_{T},\Sigma^r_{T}}$,
          \begin{align}
              &\tilde{G}_t                     =\arg\min_{G_t} V_{t+1}({\tau^a({{\Sigma^q_{T},\Sigma^1_{T},\Sigma^2_{T},\Sigma^r_{T}},G_t}})),         \nonumber \\
              &V_t({\Sigma^q_{T},\Sigma^1_{T},\Sigma^2_{T},\Sigma^r_{T}})  :=                  V_{t+1}(\tau^a({\Sigma^q_{T},\Sigma^1_{T},\Sigma^2_{T},\Sigma^r_{T}}, \tilde{G}_t)) \label{Eqn:V_subs}
          \end{align} 
\end{enumerate}

\subsection{A dynamic program for total power constraint}
In this section, we assume that both the sender and the receiver have total power constraints such that $\sum_{t=1}^T var(X_t^i)=\sum_{t=1}^TP_t^i\leq P^i, i=1,2$. 

Let $\xi_t^i$ represents total remaining power budget of player $i$ at time $t$, where $\xi_1^i = P^i$ and $\xi_{t+1}^i = \xi_t^i-P_t^i$ and $P_t^i\leq \xi_t^i,$ $\forall t,i$. 
\eq{
&(\Sigma^q_{t+1},\Sigma^2_{t+1|t+1},\xi_{t+1}^1,\xi_{t+1}^2) \nn\\
&= \tau^2(\Sigma^q_{t},\Sigma^2_{t|t},\xi_t^1,\xi_t^2,G_t,\phi_t^1,\phi^2_t).
}
where $\tau^2$ is defined through Section III and updates of $\xi_t^i$

Based on this, we propose a \emph{dynamic program} to compute optimal linear policies as follows.
\begin{enumerate}
    \item $\forall$ $\Sigma^q_{T,\Sigma^1_{T}}$, $\Sigma^2_{T},\Sigma^r_{T}$,
            $  V_{T}(\Sigma^q_{T},\Sigma^1_{T},\Sigma^2_{T},\Sigma^r_{T}, \xi_T^1, \xi_T^2) = (\Sigma^r_{T}(1,1))^2+ (\Sigma^r_{T}(2,2))^2 $.
    \item For $t=T-1,\ldots,1$, $\forall$ $\Sigma^q_{t}$, $\Sigma^2_{t},\xi_t^1,\xi_t^2$,
          \begin{align}
              &\tilde{G}_t,\tilde{\phi}_t                     =\arg\min_{G_t}  V_{t+1}({\tau({\Sigma^q_{t},\Sigma_t^1,\Sigma^2_{t},\Sigma_t^r,\xi_t^1,\xi_t^2,G_t,\phi_t}})),         \nonumber \\
              &\hspace{-1cm}V_t({\Sigma^q_{t},\Sigma^1_{t},\Sigma^2_{t},\Sigma^r_{t}},\xi_t^1,\xi_t^2)  =                  V_{t+1}(\tau(\Sigma^q_{t},\Sigma^1_{t},\Sigma^2_{t},\Sigma^r_{t}, \xi_t^1,\xi_t^2,\tilde{G}_t,\tilde{\phi}_t)) \label{Eqn:V_subs}
          \end{align} 
\end{enumerate}

\section{Conclusion}
In this paper, we consider a multiple access channel with additive white Gaussian Noise and
with active AWGN noisy feedback. We use an idea of auxiliary linear controllers at both the senders
and the receiver, as introduced in~\cite{Va20state}. Due to linearity of the policies and the controllers, all the
random variables involved are jointly Gaussian. Moreover, the corresponding covariance matrix
at the receiver of the estimation process is a deterministic process, which is a function of the
parameters of the controllers at the senders and the strategies of the players, and is thus perfectly observed by
the senders. Based on this, we formulate a dynamic program to find optimal linear policies of
the senders and the receiver that minimize the mean square probability of error at the receiver.
This is an instance of decentralized control with no common information and is one of the very
few results in the literature where such a sequential decomposition is possible.

\bibliographystyle{IEEEtran}

\end{document}